\documentclass[aps,nofootinbib,superscriptaddress,notitlepage,twocolumn]{revtex4-1}
\usepackage{epsfig}
\usepackage{amssymb,amsmath,bm,bbm,mathrsfs}
\usepackage{graphicx}
\usepackage{verbatim}
\usepackage{simplewick}
\usepackage{appendix}
\usepackage[usenames,dvipsnames]{xcolor}
\def\dd{\mathrm{d}}
\def\Mpl{M_{\rm Pl}}
\newcommand{\beal}{\begin{equation}\begin{aligned}}
\newcommand{\enal}{\end{aligned}\end{equation}}

\begin{document}

\title{Implications of Gravitational-wave Production from Dark Photon Resonance to Pulsar-timing Observations and Effective Number of Relativistic Species}

\author{Ryo Namba}
\email{ryo\_namba@sjtu.edu.cn}
\affiliation{Tsung-Dao Lee Institute, Shanghai Jiao Tong University, Shanghai 200240, China}

\author{Motoo Suzuki}
\email{m0t@icrr.u-tokyo.ac.jp}
\affiliation{Tsung-Dao Lee Institute, Shanghai Jiao Tong University, Shanghai 200240, China}

\date{\today}

\begin{abstract}
The coherent oscillation of axionic fields naturally drives copious production of dark photon particles in the early universe, due to resonance and tachyonic enhancement. During the process, energy is abruptly transferred from the former to the latter, sourcing gravitational wave generation. The resulting gravitational waves are eventually to be observed as stochastic background today. We report analytical results of this production and connect them to the recent pulsar-timing results by the NANOGrav collaboration. We show an available parameter space, around the mass $m_\phi \sim 10^{-13} \, {\rm eV}$ and the decay constant $f_\phi \sim 10^{16} \, {\rm GeV}$ with a dimensionless coupling of ${\cal O}(1)$, for our mechanism to account for the signal. 
A mechanism to avoid the axion over-dominating the universe is a necessary ingredient of this model, and we discuss a possibility to recover a symmetry and render the axion massless after the production.
We also comment on potential implications of the required effective number of relativistic species to the determination of the present Hubble constant.
\end{abstract}

\maketitle

\section{Introduction}
\label{sec:intro}
Pseudo-Nambu-Goldstone bosons arise from spontaneous breaking of global symmetries and are ubiquitous in UV complete theories beyond the Standard Model (SM). They may serve as a solution to the strong CP problem by the Peccei-Quinn mechanism (QCD axion) \cite{Peccei:1977hh,Peccei:1977ur,Weinberg:1977ma,Wilczek:1977pj} and/or may act as dark matter \cite{Abbott:1982af,Dine:1982ah,Preskill:1982cy,Ipser:1983mw}.
In this sense, they bridge between the fundamental theories beyond SM and low-energy observables. Hereafter, we refer to them as axion-like fields (ALFs).

An intriguing nature of ALF $\phi$ is their unique coupling to other field contents. In particular, its coupling to a $U(1)$ gauge field
\begin{equation}
    {\cal L}_{\rm int} = - \frac{\alpha}{4 f_\phi} \, \phi \, F_{\mu\nu} \tilde{F}^{\mu\nu} \; ,
    \label{Lint}
\end{equation}
is generically allowed.
Here $F$ and $\tilde{F}$ are the field-strength tensor of the gauge field $A_{\mu}$ and its dual, respectively, $f_\phi$ is a constant of mass dimension one, sometimes called axion decay constant, and $\alpha$ is a dimensionless constant.
If $\phi = {\rm const.}$, the term \eqref{Lint} is topological and has no effect on the dynamics of the system, at least perturbatively. In other words, one can rewrite ${\cal L}_{\rm int} = \frac{\alpha}{2f_\phi} \, \partial_\mu \phi \, A_\nu F^{\mu\nu} $ up to total derivatives and would be vanishing if $\partial \phi =0$. This observation implies that \eqref{Lint} is indeed compatible with the axion's intrinsic shift symmetry, and thus should be included in models of $\phi$ in the language of effective field theory (EFT).
In this paper we stay agnostic about the identity of $A_\mu $ and refer to it by ``dark photon.''

Phenomenology of the coupling \eqref{Lint} in cosmological settings has been extensively studied in the past years, such as
inflationary model buildings \cite{Anber:2009ua,Peloso:2015dsa,Notari:2016npn,Ferreira:2017lnd,Tangarife:2017vnd,Tangarife:2017rgl,Almeida:2018pir},
cosmic microwave background (CMB) observables \cite{Lue:1998mq,Barnaby:2010vf,Sorbo:2011rz,Barnaby:2011vw,Barnaby:2011pe,Dimopoulos:2012av,Anber:2012du,Meerburg:2012id,Linde:2012bt,Ferreira:2014zia,Bartolo:2015dga,Ferreira:2015omg,Peloso:2016gqs,Alexander:2017bxe,Domcke:2018eki,Almeida:2019hhx,Domcke:2019qmm,Domcke:2020zez},
generation of magnetic fields \cite{Durrer:2010mq,Ng:2015ewp,Fujita:2015iga,Adshead:2016iae,Caprini:2017vnn,Shtanov:2019civ,Shtanov:2019gpx,Patel:2019isj,Fujita:2019pmi,Sobol:2019xls},
formation of primordial black holes \cite{Bugaev:2013fya,Erfani:2015rqv,Domcke:2017fix,Cheng:2018yyr,Ozsoy:2020kat},
generation of baryon asymmetry \cite{Jimenez:2017cdr},
dark matter physics \cite{Kamada:2017cpk,Agrawal:2017eqm,Co:2018lka,Bastero-Gil:2018uel,Agrawal:2018vin,Machado:2018nqk},
and non-Abelian extensions \cite{Adshead:2012kp,Adshead:2012qe,Dimastrogiovanni:2012st,Dimastrogiovanni:2012ew,Maleknejad:2012dt,Adshead:2013qp,Adshead:2013nka,Namba:2013kia,Maleknejad:2013npa,Obata:2014loa,Maleknejad:2016qjz,Dimastrogiovanni:2016fuu,Agrawal:2017awz,Thorne:2017jft,Fujita:2017jwq,Agrawal:2018mrg,Fujita:2018ndp,Papageorgiou:2018rfx,Domcke:2018rvv,Fujita:2018vmv,Dimastrogiovanni:2018xnn,Papageorgiou:2019ecb,Watanabe:2020ctz,Almeida:2020kaq}.
Some of these models have been directly tested by the Planck mission \cite{Ade:2013ydc,Ade:2015lrj,Ade:2015ava,Akrami:2019izv}.
The interaction \eqref{Lint} induces a copious production of gauge quanta in the presence of coherent motion of $\phi$ \cite{Anber:2009ua}, resulting in various observational signals. Our focal point of interest in this paper is generation of gravitational waves (GWs) sourced by such produced gauge fields, or dark photons. In this context, past studies have been performed for GWs as the CMB tensor modes \cite{Barnaby:2012xt,Cook:2013xea,Shiraishi:2013kxa,Mukohyama:2014gba,Mirbabayi:2014jqa,Namba:2015gja,Domcke:2016bkh,Shiraishi:2016yun,Obata:2016oym,Ozsoy:2020ccy} as well as GW signals at terrestrial interferometers \cite{Cook:2011hg,Barnaby:2011qe,Crowder:2012ik,Garcia-Bellido:2016dkw,Obata:2016tmo,Obata:2016xcr,Machado:2019xuc,Okano:2020uyr}, and future observational prospects have been discussed for LiteBIRD \cite{Shiraishi:2016yun} and for LISA \cite{Bartolo:2016ami}.

Once the axion mass overcomes the Hubble friction, $\phi$ starts oscillating coherently at some moment in the cosmic history. This oscillation can trigger a resonant amplification of the dark photon, together with a tachyonic enhancement for a certain fraction of each oscillation in the cases of large coupling. The growth of this type from the interaction \eqref{Lint} has been studied in the literature, \cite{Adshead:2015pva,McDonough:2016xvu,Cuissa:2018oiw,Cheng:2015oqa,Kitajima:2017peg} for the amplification mechanism itself, and \cite{Adshead:2018doq,Adshead:2019igv,Adshead:2019lbr,Salehian:2020dsf} for its contribution to GW signals.
All of these works are based on numerical methods including lattice simulation, with the only exception by \cite{Salehian:2020dsf}, in which analytical results are given with the main focus on a large coupling case. Our analysis in this paper utilizes the analytical calculations we have conducted independently and cross-checked with those in \cite{Salehian:2020dsf}. The details of our calculations will be discussed in our upcoming publication \cite{NS_axion}, and the present paper is devoted to collecting the results of interest in light of the recent report of a stochastic GW signal by a pulsar-timing array (PTA) experiment.

The North American Nanohertz Observatory for Gravitational Waves (NANOGrav) \cite{McLaughlin:2013ira,Brazier:2019mmu} has found a significant Bayes factor in favor of the presence of stochastic GW background in their $12.5$-year PTA data \cite{Arzoumanian:2020vkk}. Their current result shows no statistically significant evidence for the presence of quadrupolar spatial correlations and thus cannot conclude a definitive detection of GW background that is consistent with the General Relativity (GR). It may have been caused by spin noise, solar system effects, or other unknown systematics, and disentanglement of these systematics from the true signals needs to await further analyses and data from the other PTA experiments \cite{Arzoumanian:2020vkk}.
 Nevertheless, other possibilities are worth exploring, assuming that the NANOGrav 12.5-year signal result from a true GW back- ground of astrophysical or cosmological origin.
Possible sources of stochastic GW signals include mergers of supermassive black hole binaries \cite{Rajagopal:1994zj,Huerta:2015pva,Jaffe:2002rt,Wyithe:2002ep,1818890}, cosmic string network in the early universe \cite{Ellis:2020ena,Blasi:2020mfx,Buchmuller:2020lbh,Samanta:2020cdk} (see e.g.~\cite{Vilenkin:1981bx,Vachaspati:1984gt,Ringeval:2005kr,Siemens:2006yp,Kawasaki:2010yi,BlancoPillado:2011dq,Blanco-Pillado:2017rnf,Ringeval:2017eww} for earlier works), oscillating GW sound speed \cite{Cai:2020ovp}, fast radio burst sources \cite{Pearson:2020wxb}, blue spectrum of the inflationary tensor mode \cite{Vagnozzi:2020gtf}, and primordial black holes \cite{Vaskonen:2020lbd,DeLuca:2020agl,Bhaumik:2020dor,Kohri:2020qqd}. Phase transitions in the early universe have been actively investigated as a GW source \cite{Randall:2006py,Caprini:2010xv,Schwaller:2015tja,Jaeckel:2016jlh,Kobakhidze:2017mru,Iso:2017uuu,Breitbach:2018ddu,Fujikura:2018duw,Baratella:2018pxi,Megias:2018sxv,Agashe:2019lhy,Fujikura:2019oyi,DelleRose:2019pgi,vonHarling:2019gme,Ghoshal:2020vud} and considered in the context of the NANOGrav result in \cite{Nakai:2020oit,Addazi:2020zcj}.

In this paper, we explore the dynamics of interacting ALF and dark photon as a source of stochastic GW signals. Once the axion starts oscillating coherently due to its own mass, the dark photon is significantly amplified due to resonance with the axion and tachyonic instability. If this occurs when the cosmic temperature is $T \lesssim 0.1 \, {\rm GeV} $, a GW spectrum that covers the frequency range of the NANOGrav signals can be achieved. Due to this process, the axion energy is efficiently transferred to the dark photon, and the amplitudes of the observed signals are reached as long as the coupling is strong enough to draw the sufficient energy out of the axion.
This requires rather a large energy content of the axion to produce a sufficient level of GW background. We discuss a possible mechanism to render the axion massless after the dark photon production and also consider the effective number of relativistic species prior to the recombination. We comment on the implication of this requirement on the determination of the present value of the Hubble parameter and on its potential alleviation of the tension in the measurements of $H_0$~\cite{Bernal:2016gxb,Aghanim:2018eyx,Blinov:2020hmc}.

The rest of this paper is organized as follows. We set up the simple model of our interest in Sec.~\ref{sec:model}. In Sec.~\ref{sec:production}, we calculate the dark photon production. We first comment on the absence of production in the small coupling regime on a cosmological background spacetime, and then derive analytic expressions of the production in the case of large coupling strength. In Sec.~\ref{sec:GW}, we compute the GW spectrum induced by the produced dark photon and discuss its relevance to the NANOGrav observation. Sec.~\ref{sec:conclusion} is devoted to discussions and conclusion.
Throughout the paper, we use the natural units $\hbar = c = k_B = 1 $, denote the reduced Planck mass by $\Mpl $, and take the flat Friedmann-Lema\^{i}tre-Robertson-Walker metric as the cosmological background spacetime.

\section{Model Setup}
\label{sec:model}

An axion-like field (ALF) $\phi$ emerges from spontaneous breaking of a global symmetry characterized by an energy scale $f_\phi$. Their shift symmetry is softly broken by non-perturbative dynamics at another energy $\Lambda$. Then their mass is typically of order $m_\phi \sim \Lambda^2 / f_\phi $, whose stability against quantum corrections, necessarily proportional to $1/f_\phi$, is technically natural.
After being produced, at some point in the history of the universe, axion starts coherent oscillation within a coherent length $L_c $. Inside this region, the spatial gradient of $\phi$ is negligible, and its oscillation in a temporal direction is well approximated by
\begin{equation}
    \phi(t) \cong \phi_{\rm osc} \left( \frac{a_{\rm osc}}{a} \right)^{3/2} \cos m_\phi \left( t - t_{\rm osc} \right) \; ,
    \label{phisol}
\end{equation}
where $t$ is the cosmic time, $a$ is the cosmic scale factor, and subscript ``osc'' denotes values at the time of the onset of the coherent oscillation.

Another key feature of axion is that its shift symmetry uniquely determines the lowest-order coupling to other fields. In particular, a dark photon field $A_\mu $ that possesses a $U(1)$ gauge symmetry interacts with ALF through the term \eqref{Lint}.
The dark photon may acquire mass $m_{\gamma'}$ by Higgs-like or Stueckelberg-type mechanism~\cite{Stueckelberg:1900zz,Delbourgo:1986wz}, but a large mass would disrupt the effect of the interaction \eqref{Lint}. Hence we are interested in the parameter range where such a disturbance is absent. This requires the mass to be smaller than the coupling strength, yielding a condition $m_{\gamma'}^2 \ll k \alpha \dot\phi / f_\phi $, where dot denotes derivative with respect to $t$, and $k$ is the typical momentum of the dark photon.
Non-perturbative effects of \eqref{Lint} on the dark photon can be partially captured by solving the equation of motion for $A_\mu $.
Projecting $A_\mu$ onto circular polarization states $\hat{A}_\pm $ in the Fourier space, the E.o.M.~of the latter reads, under a negligible dark photon mass,
\begin{equation}
    \left( \partial_\tau^2 + k^2 \mp k \, \frac{\alpha}{f_\phi} \, \partial_\tau \phi \right) \hat{A}_\pm = 0 \; ,
    \label{EOM_A}
\end{equation}
where $\tau$ is the conformal time $\dd \tau = \dd t / a$.
Inside the region of the coherent oscillation \eqref{phisol}, the dispersion relation of $\hat{A}_{\pm}$ in the coordinates of the physical time, defined by $\omega_\pm^2 \equiv k^2/a^2 \mp k \alpha \dot\phi / (a f_\phi) $, is approximately
\begin{equation}
    \omega_\pm^2 \cong \frac{k^2}{a^2} \pm m_\phi \, \frac{k}{a} \, \frac{\alpha \phi_{\rm osc}}{f_\phi} \left( \frac{a_{\rm osc}}{a} \right)^{3/2} \sin m_\phi \left( t - t_{\rm osc} \right) \; .
    \label{dispersion}
\end{equation}
Without the cosmic expansion $a = {\rm const.}$, \eqref{EOM_A} with \eqref{dispersion} would yield the Mathieu equation (see e.g.~\cite{McLachlan} for a detailed analysis). In this work, we include the effect of the expansion and derive analytical expressions, to attempt explaining the recent result of NANOGrav.

\section{Dark photon production}
\label{sec:production}

In a Minkowski spacetime, the equation \eqref{EOM_A} with the dispersion \eqref{dispersion} would be of the form of the Mathieu equation, and the dark photon field would resonate with the oscillating axion. If the amplitude of the oscillation were small, so-called \textit{narrow resonance} would take place, and only some limited momentum/frequency bands would be enhanced. For a large oscillation amplitude, on the other hand, a much wider range of momentum values would get resonated, called \textit{broad resonance}. See \cite{McLachlan,Kofman:1997yn,Greene:1997fu} for details.

The structure of resonance is, however, modified in the more realistic, expanding universe. 
The modification is not only quantitative, but arises already at a qualitative level
\cite{Kofman:1997yn}. In particular, would-be narrow resonance bands are no longer available if the expansion is taken into account, and thus there is no amplification of dark photon for a small ALF amplitude. The condition for this case can be quantified by an upper bound on the coupling strength,
\begin{equation}
    \frac{\alpha \phi_{\rm osc}}{f_\phi} < \frac{k}{a_{\rm osc} m_\phi} \left( \frac{a}{a_{\rm osc}} \right)^{1/2} \; , \quad
    \mbox{small coupling} \; .
    \label{narrowband}
\end{equation}
The absence of narrow resonance can be understood as follows: for this type of resonance, only a limited range of modes $k$ range would grow. In a flat spacetime, the primary band width of the resonance in our model \eqref{dispersion} could be quantified by $\vert k - m_\phi /2 \vert \lesssim \alpha \phi_{\rm osc} m_\phi / f_\phi $. Outside of this small window, no resonance would take place. Note that, because of this narrow band $k \approx m_\phi /2$, we observe from \eqref{narrowband} that $\alpha \phi_{\rm osc} / f_\phi < 1 $ for a narrow resonance in a flat spacetime.
The formal reason of the primary-band growth is that the oscillation of $\hat{A}_\pm $ due to the matching momentum $k$ should be canceled out by the ALF's oscillation due to its mass $m_\phi$, and this non-oscillatory piece would be the one that grows.
However, the expansion of space changes the physical momentum by $k / a$ over time, which completely alters the nature of the resonance. While it is crucial for the momentum to stay in the resonance band during the time scale of the growth, the expansion only allows the cancellation between $k/a$ and $m_\phi$ to last for a short duration of $k \Delta \tau \sim {\cal O}(1) $, where $\tau$ is the conformal time. Around a would-be resonating momentum $k/a \sim m_\phi$, this corresponds to only a few oscillations. In the regime of narrow resonance $\alpha \phi_{\rm osc} / f_\phi $, this does not provide sufficient time for the mode to grow. After this duration, the cancellation ceases, and no further growth is expected.
One might still suspect that, even if \textit{each mode} did not grow sufficiently, a \textit{collection} of small amplifications of different modes would contribute to a large value, since different $k$ values would equate $m_\phi $ at different times due to the expansion. This turns out not to be the case, and every mode simply experiences no amplification, and integration over $k$ is no different from the case of no resonance from the start.
In the following subsection, we therefore concentrate on studying the case of large coupling strength.
The statements in this paragraph, as well as the following calculations of the productions, will be discussed in detail in our upcoming work \cite{NS_axion}.

\subsection{Large coupling}
\label{subsec:large coupling}

The range of large coupling is the regime opposite to \eqref{narrowband}, i.e.,
\begin{equation}
    \frac{\alpha \phi_{\rm osc}}{f_\phi} > \frac{k}{a_{\rm osc} m_\phi} \left( \frac{a}{a_{\rm osc}} \right)^{1/2} \; , \quad
    \mbox{large coupling} \; .
    \label{broadband}
\end{equation}
There are $2$ physical mechanisms of copious particle production that are in action: growth by tachyonic instability, and violation of adiabaticity. Both of these two effects occur for a given mode, but at different moments, and repeat as long as the axion oscillation continues. 
A necessary condition leading to the tachyonic instability is given as
\begin{align}
\label{eq:tachyonic}
\omega^2_{\pm}<0\ .
\end{align}
The adiabaticity of the system is characterized by a quantity $\left|\dot{\omega}_{\pm}/\omega^2_{\pm}\right|$, and the adiabatic condition is violated in the region of 
\begin{align}
\left|\frac{\dot{\omega}_{\pm}}{\omega^2_{\pm}}\right|\gtrsim 1\ .
\end{align}
We solve the field equation of motions in each region in analytical way, and connect the solutions step by step. After a straightforward calculation, the exponential growth factor of the gauge field mode functions $A_\pm$ is found to be
\begin{align}
\label{eq:mu_m}
&\ln(|A_\pm|)\equiv \mu^{\pm}_m\simeq(m-2)\log(2)\nonumber\\
&+\tilde\gamma\left[
   \left(m+\frac{m_\phi t_{\rm osc}}{2 \pi
   }-\frac{3}{4}\pm\frac{1}{4}\right)^{\frac{1 + 6w}{6(1+w)}}
- \left(\frac{m_\phi t_{\rm osc}}{2 \pi}+\frac{1}{4}\pm\frac{1}{4}\right)^{\frac{1 + 6w}{6(1+w)}}
\right] \ .
\end{align}
Here, $w$ is the equation of state of the universe, $m$ is an integer $m=2,3,..$ denoting the $m$-th cycle of the axion oscillation,%
\footnote{That is, the time $t$ within the $m$-th cycle spans the range $m_\phi t_{\rm osc}+2\pi(m-1) \le m_\phi t < m_\phi t_{\rm osc}+2\pi\,m$.}
the initial amplitude $A_{\pm}$ at $t=t_{\rm osc}$ is normalized as $1$, and the factor $\tilde\gamma$ is given as
\begin{align}
\tilde\gamma \equiv &
\frac{2^{\frac{5(2 + 3w)}{6(1+w)}} 3 \pi ^{-\frac{8 + 3w}{6 (1+w)}} ( 1 + w ) \, \Gamma \left(\frac{3}{4}\right)^2} 
   {1 + 6w}
   \nonumber\\ & \times
   \left( m_\phi\,t_{\rm osc} \right)^{\frac{5}{6(1+w)}}
   \sqrt{\left(\frac{k}{m_\phi\,a_{\rm osc}}\right)\frac{\alpha \phi_{\rm osc}}{f_\phi}} \ ,
\label{gamma}
\end{align}
where $a_{\rm osc}$ is the value of the scale factor at $t=t_{\rm osc}$. The first term in Eq.\,\eqref{eq:mu_m} comes from the adiabaticity violation and the second term is obtained from the tachyonic instability. 
Note that the premise in obtaining the expressions for $\mu_m^\pm$ in \eqref{eq:mu_m} is that the coupling strength is large. As can be speculated from \eqref{broadband}, this ``large coupling limit'' is in fact the leading-order expression in the expansion in terms of the small parameter $\frac{k/(a_{\rm osc}m_\phi)}{\alpha\,\phi_{\rm osc}/f_\phi} \ll 1$. Indeed, if one included subleading-order terms, they would be suppressed by this parameter compared to the term in \eqref{eq:mu_m} \cite{Salehian:2020dsf}.
We discuss the validity of this approximation in the next section.

%%%%
\begin{figure}
\centering
  \includegraphics[width=0.9\linewidth]{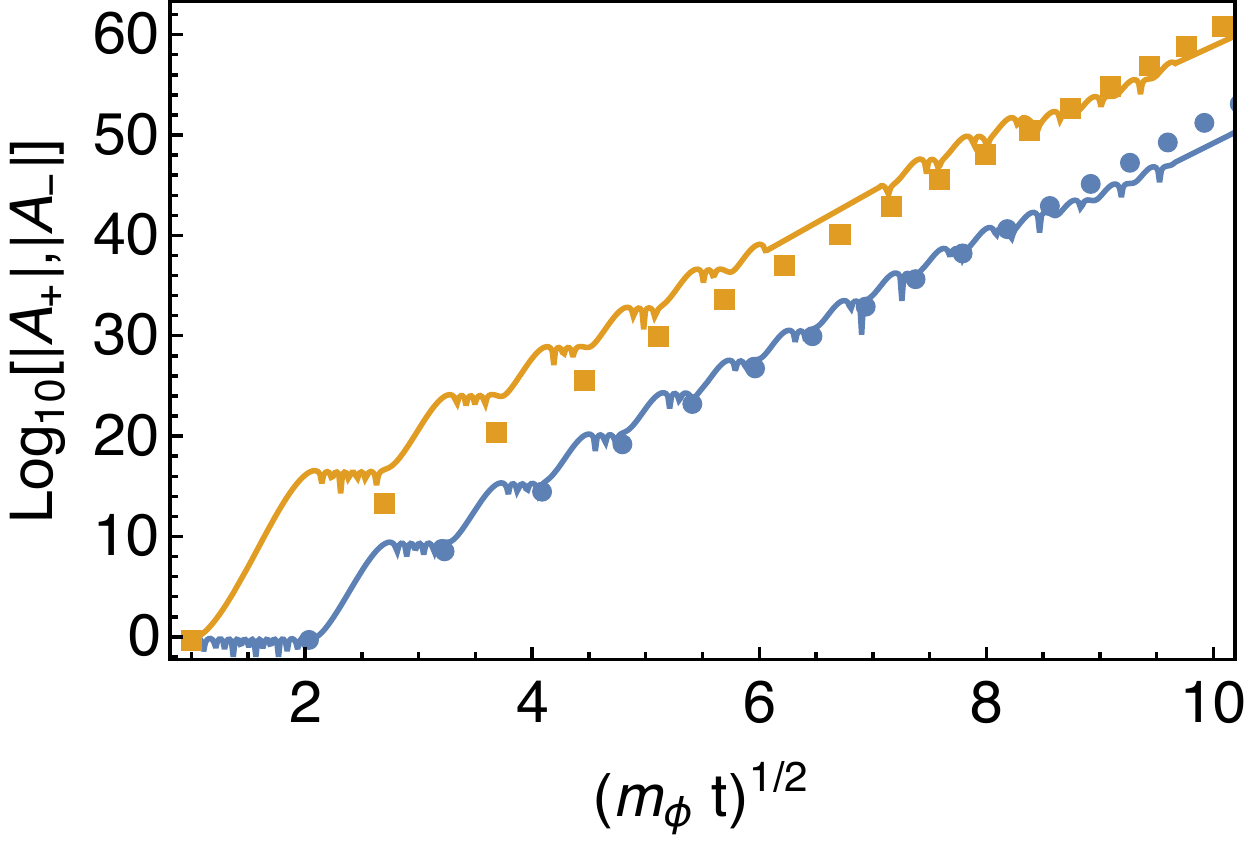}
   \caption{Comparing the analytical and numerical calculation for $A_\pm$. The yellow square and blue circle denote the analytical results for $A_-$ and $A_+$, respectively. The yellow and blue solid lines are the numerical results of $A_-$ and $A_+$, respectively. Here, we take the parameters as $w=1/3$, $m_\phi \,t_{\rm osc}=1$, $k\,t_{\rm osc} / a_{\rm osc} =0.5$, $\alpha\phi_{\rm osc} / f_\phi =1.5\times10^{3}$. We also normalized the initial amplitude as $A_{\pm}(t_{\rm osc})=1$.
   }
\label{fig:mu_m}
\end{figure}
%%%%  

In Fig.~\ref{fig:mu_m}, we compare our analytical results \eqref{eq:mu_m} to the numerical computation.
We take the parameters as $w=1/3$ (radiation domination), $m_\phi \,t_{\rm osc}=1$, $k\,t_{\rm osc} / a_{\rm osc}=0.5$,
%($=\kappa/2$ in my notation)
$\alpha\phi_{\rm osc} / f_\phi =1.5\times10^{3}$,
%($=Q/2$ in my notation)
and $A_{\pm}(t_{\rm osc})=1$. The yellow squares and blue circles indicate the analytical results of $A_-$ and $A_+$, respectively, evaluated at the end of the flat region of each cycle. The solid yellow and blue lines, respectively, correspond to the numerically computed amplitudes of $A_-$ and $A_+$. Here, the growth appears to continue indefinitely only because we do not include the back reaction effects. We confirm a nice agreement between the analytical and numerical calculations.
In more details, in the oscillating but flat amplitude regions in Fig.~\ref{fig:mu_m}, the adiabatic condition is not violated, nor does tachyonic instability takes place, and therefore no gauge field is produced.  The growing regions correspond to the periods where tachyonic instability occurs. Adiabaticity condition is violated in the regions sandwiched between the former two regions. Note that the time evolution of $A_{\pm}$ are different because the tachyonic instability condition \eqref{eq:tachyonic} is satisfied at different timings for $\omega_{\pm}$. 
This is due to the phase difference appearing in \eqref{dispersion} as the $\pm$ sign, resulting from the parity breaking interaction \eqref{Lint} in the presence of nonzero $\dot\phi$.
For the consideration in the following sections, we concentrate on the dark photon production during the era of radiation domination, and thus we set $w = 1/3$ from here on.

\section{Gravitational waves and NANOGrav results}
\label{sec:GW}

We now turn to the estimation of the gravitational-wave (GW) generation sourced by the produced dark photon computed in Sec.~\ref{subsec:large coupling}. 
GW represents the pure gravitational degrees of freedom that propagate in vacuum and can be identified with the traceless and transverse part of the metric perturbations, that is
$ h_{ij}\equiv a^{-2}\delta g_{ij} $, with the properties $\partial_i h_{ij} = h_{ii} = h_{[ij]} = 0 $. The sourced contribution to GW from the dark photon is computed from the traceless and transverse part of the Einstein equations.
Projected onto the polarization states $\hat{h}_\lambda (\tau, \bm{k}) $ along the wavenumber $\bm{k}$ in the Fourier space, these equations read
\begin{equation}
    \left( \partial_\tau^2 + k^2 - \frac{\partial_\tau^2 a}{a} \right) \left( a \hat{h}_\lambda \right) = \hat J_\lambda (\tau, \bm{k}) \; ,
    \label{EOM_GW}
\end{equation}
where $k \equiv \vert \bm{k} \vert$ and
\begin{equation}
    \hat J_\lambda = \frac{2 a}{\Mpl^2} \, \Pi_\lambda^{ij}(\hat{k}) \int \frac{\dd^3 x}{(2 \pi)^{3/2}} \, {\rm e}^{- i \bm{k} \cdot \bm{x}} \, T_{ij} (\tau , \bm{x}) \; ,
\end{equation}
$\Mpl$ denotes the reduced Planck mass, $\tau$ the conformal time, and $\Pi_\lambda^{ij}(\hat{k}) $ the inverse of the GW polarization tensor. Here the traceless and transverse part of $T_{ij}$ is projected by multiplying $\Pi_\lambda^{ij} $. The Green function for $a \hat{h}_\lambda $ is found to be, inside the Hubble horizon,
\begin{equation}
    G_k (\tau , \tau') = \Theta( \tau - \tau' ) \, \frac{\pi}{2} \sqrt{\tau \tau'} \left[ Y_\nu ( k \tau) \, J_\nu ( k \tau' ) - J_\nu ( k \tau ) \, Y_\nu ( k \tau' ) \right] \; ,
\end{equation}
where $\Theta(x)$ is the Heaviside step function, $J_\nu(x) $ and $Y_\nu(x) $ are the Bessel functions of the first and second kinds, respectively, with the index $\nu = 3 (1 - w) / 2(1 + 3w) $ for the equation of state $w \in ( -1/3, 1)$, and thus $\nu = 1/2$ for radiation domination $w = 1/3$. For small wavelength modes that satisfy $k^2 \gg \partial_\tau^2 a / a$, the Green function is approximated to be simply $G_k(\tau , \tau') \simeq \Theta( \tau - \tau' ) \, k^{-1} \sin k ( \tau - \tau' ) $.
Then the particular solution of \eqref{EOM_GW} sourced by $J_\lambda $ is obtained by the Green function method as
\begin{equation}
    \hat{h}_\lambda (\tau, \bm{k}) = \frac{1}{a(\tau)} \int_{-\infty}^{\infty} \dd\tau' \, G_k(\tau, \tau') \, \hat J_\lambda( \tau', \bm{k} ) \; .
\label{GreenMethod}
\end{equation}
The associated GW energy density $\rho_{\rm GW} $ is
\begin{equation}
    \rho_{\rm GW} \equiv \frac{\Mpl^2}{8 a^2} \, \langle \partial_\tau h_{ij} \, \partial_\tau h_{ij} + \partial_k h_{ij} \, \partial_k h_{ij} \rangle \; ,
    \label{rhoGW}
\end{equation}
where $\langle \bullet \rangle$ denotes the spatial average, and the GW fields are assumed to vanish at the spatial infinity.

To compare with the pulsar-timing data in \cite{Arzoumanian:2020vkk}, it is convenient to compute the spectrum of the fractional GW energy density, defined by
\begin{equation}
    \Omega_{\rm GW, 0} \equiv \frac{1}{3 H_0^2 \Mpl^2} \, \frac{\dd \rho_{\rm GW}(t_0)}{\dd \ln k} \; ,
\end{equation}
evaluated at the present time $t = t_0$.
To connect this value $\Omega_{\rm GW, 0}$ to the value at the time of generation, denoted by $\Omega_{\rm GW, gen} $, we assume the entropy conservation, $3$ neutrino species, free propagation of GW after production ends, and that the GW value is averaged over oscillations. Then we find \cite{Boyle:2005se},
\begin{equation}
\begin{aligned}
    \Omega_{{\rm GW},0} &
    \approx 0.32
    \left( \frac{g_{s, {\rm 0}}}{g_{s, {\rm gen}}}\right)^{4/3} \frac{g_{*,{\rm gen}}}{g_{*,{\rm 0}}} \,
    \Omega_{r,0} \,
    \Omega_{\rm GW, gen} \; ,
\end{aligned}
\label{OmegaGW_transfer}
\end{equation}
where $g_{*,{\rm gen}}$ and $g_{s,{\rm gen}} $ are the number of relativistic degrees of freedom for the energy density and entropy, respectively, at the time of production, and $\Omega_{r,0}h^2\simeq 4.16\times 10^{-5}$ with $h \approx 0.67$ is the current value of the fractional radiation density \cite{Aghanim:2018eyx}. Thus, once we find the spectrum of GW at the production using \eqref{rhoGW}, the corresponding value at present is trivially obtained by \eqref{OmegaGW_transfer}.

Using the result for the dark photon production obtained in  Sec.~\ref{subsec:large coupling}, and using \eqref{GreenMethod} and \eqref{rhoGW}, we find the spectrum of the GW energy density at the time of generation is \cite{Salehian:2020dsf}
\begin{align}
\label{eq:gw_spectrum}
&\left.\Omega_{\rm GW, \gamma'}\right\vert_{\rm gen}\approx
\frac{n_{\rm gen}^2 H_{\rm osc}k_s^9}{96\pi^3 M_{\rm PL}^4m_\phi^4H_{\rm gen}^2 a_{\rm gen}^4 a_{\rm osc}^5}\left(\frac{f_\phi}{\alpha \phi_{\rm osc}}\right)^2
\nonumber \\
& \quad \times
\left(\frac{k}{2k_s}\right)\left(1-\frac{k^2}{4k_s^2}\right)^3\left[\left(1-\frac{k}{2k_s}\right)^4+\left(1+\frac{k}{2k_s}\right)^4\right]\ ,
\end{align}
where $\Omega_{\rm GW, \gamma'}$ denotes the fractional density of GW sourced by the dark photon, and the subscript ``gen'' indicates the generation time of GW, 
$k_s$ is the wavenumber of the dominant growth mode of the photon given by~\cite{Machado:2018nqk,Salehian:2020dsf}
\begin{align}
\label{eq:dominant_momentum}
\frac{k_s}{a_{\rm gen}}\approx  \frac{m_\phi}{2^{5/6}3^{1/6}}\frac{a_{\rm osc}}{a_{\rm gen}}\left(\frac{m_\phi\alpha\phi_{\rm osc}}{f_\phi \,H_{\rm osc}}\right)^{2/3} ,
\end{align}
and $n_{\rm gen}$ is the occupation number of the dark photon for the mode $k_s$. 
The gravitational wave spectrum in Eq.\,\eqref{eq:gw_spectrum} is obtained by assuming the dark photon is produced during the radiation dominated universe, and the spectrum of the produced photon has delta function-like peak at $k_s$~\cite{Salehian:2020dsf}. Furthermore, we only take into account $A_-$ mode which is the dominant mode as we have seen in Fig.~\ref{fig:mu_m}.

\begin{comment}
%%%%
\begin{figure}
\centering
  \includegraphics[width=0.9\linewidth]{back_reaction.pdf}
   \caption{Comparison of $a_{\rm br}/a_{\rm osc}$ and  $a_{\rm tac}/a_{\rm osc}$ as a function of the coupling strength $\alpha \phi_{\rm osc} / f_\phi$. The blue dashed curve corresponds to $a_{\rm br}/a_{\rm osc}$ for $m_\phi=2H_{\rm osc}$, while the red solid curve to $a_{\rm tac}/a_{\rm osc}$. From the figure, we observe that it is consistent to use $\alpha \phi_{\rm osc}/f_\phi\approx30$ and $a_{\rm br}\approx a_{\rm tac}\approx 10$ in our analysis.
   }
\label{fig:back_reaction}
\end{figure}
%%%%  
\end{comment}

So far we have assumed that the resonant production of the photon continues as long as the tachyonic instability condition in Eq.\,\eqref{eq:tachyonic} is satisfied. This assumption is not suitable once the back reaction effects become substantial, since they are expected to disturb the resonance. The time $a_{\rm br}$ when the back reaction stops the resonance is estimated by comparing the terms $m_\phi^2\phi$ and $\frac{\alpha}{4f}F_{\mu\nu}\tilde F^{\mu\nu}$ in the equation of motion $\phi$, i.e.
\begin{align}
\label{eq:br}
&m_\phi^2\phi_{\rm osc}\left(\frac{a_{\rm osc}}{a_{\rm br}}\right)^{3/2}\sim \frac{\alpha}{f_\phi}\frac{n_{\rm gen}k_s^4}{2\pi^2 a_{\rm br}^4}
\end{align}
where the right-hand side of the above equation is obtained by focusing on the dominant photon mode $k_s$ and taking $a_{\rm br}\sim a_{\rm gen}$. The occupation number $n_{\rm gen}$ is roughly estimated as $n_{\rm gen}\approx |A_-(k_s)|^2$ from Eq.\,\eqref{eq:mu_m}, where $A_-$ is the value of the mode function but normalized to unity at the initial time $t = t_{\rm osc}$.
On the other hand, the tacyonic instability condition \eqref{eq:tachyonic} can be met until the time $a_{\rm tac}$, which is obtained by using Eq.\,\eqref{broadband} and Eq.\,\eqref{eq:dominant_momentum},
\begin{align}
\label{eq:atac}
a_{\rm tac}\approx a_{\rm osc}\left(\frac{\alpha\phi_{\rm osc}}{f_\phi}\right)^{2/3}\ ,
\end{align}
where we have also used $m_\phi\sim H_{\rm osc}$. 
Now we can compute $a_{\rm br}$ by solving Eq.\,\eqref{eq:br} and taking $n_{\rm gen} \approx \exp(2\mu_m)$.  We note from \eqref{gamma} that $a_{\rm br}$ is sensitive to the quantity $\alpha \phi_{\rm osc}/f_\phi$, while the dependence of $m_\phi$ and $\phi_{\rm osc}$ are logarithmic and negligible for the precision of our computation.
In obtaining \eqref{eq:gw_spectrum}, we have implicitly assumed that the production ceases to operate because of the termination of the tachyonic instability. On the other hand, the produced GW abundance becomes maximum when the dark photon is produced to the extent at which it starts back-reacting to the ALF motion. Therefore, the optimal scenario for the $\Omega_{\rm GW, \gamma'}$ value within the validity range of our calculation is the case where those two moments coincide. We thus equate $a_{\rm br}$ given by \eqref{eq:br} and $a_{\rm tac}$ by \eqref{eq:atac}, yielding $\alpha \phi_{\rm osc}/f_\phi\approx30$ and $a_{\rm br}\approx a_{\rm tac}\approx 10 a_{\rm osc}$. This is our main target parameter region. 
The above condition $\alpha \phi_{\rm osc}/f_\phi\approx30$ can be satisfied, for example, by  $\phi_{\rm osc}/f_\phi\gg 1$
in the context of the clockwork mechanism~\cite{Choi:2015fiu,Kaplan:2015fuy,Giudice:2016yja,Farina:2016tgd}.

In fact, our analytical expression \eqref{eq:mu_m} is obtained in the limit of large coupling,
 i.e.~the leading-order expression in the expansion with respect to the parameter $(k/a_{\rm osc} m_\phi) (\alpha \phi_{\rm osc} / f_\phi)^{-1}$, as mentioned below Eq.\,\eqref{gamma}.
In particular, around the peak momentum $k/(a_{\rm osc} m_\phi) \sim (\alpha \phi_{\rm osc} / f_\phi)^{2/3}$, this parameter is $\propto (\alpha \phi_{\rm osc} / f_\phi)^{-1/3}$, and the expansion is not particularly accurate for our target value $\alpha \phi_{\rm osc} / f_\phi \approx {\cal O}(10)$.
This fact is potentially followed by an overestimation of $n_{\rm gen}$, and in turn the actual time of the production termination, $a_{\rm br} \sim a_{\rm tac}$, may be delayed compared to the purely analytical calculation. As we see soon below, this would not alter our conclusion regarding the GW spectrum in view of the NANOGrav data, but it would tighten the constraint on the effective number of relativistic degrees of freedom, $\Delta N_{\rm eff}$.

The prediction from our model is now to be tested against the results by NANOGrav \cite{Arzoumanian:2020vkk}. This observation evaluates $\Omega_{{\rm GW},0} $ as a function of frequency $f$ in the form \cite{Arzoumanian:2018saf},
\begin{equation}
    \Omega_{\rm GW,0}(f) =  \frac{2\pi^2 f_{\rm yr}^2}{3 H_0^2} \, \left( \frac{f}{f_{\rm yr}} \right)^{5-\gamma}
    A_{\rm GWB}^2
    \label{OmegaGW_NANOGrav}
\end{equation}
where $A_{\rm GWB}$ is the amplitude of the gravitational wave of an assumed power-law spectrum with a spectral index $\gamma$, $f_{\rm yr}=1\,{\rm yr}^{-1}$, and $H_0$ is the Hubble parameter at present. We are particularly interested in fitting the spectrum
$\Omega_{\rm GW,0}(f)$
by the power law $\gamma=4$, since our GW spectrum is proportional in $f$ as in Eq.\,\eqref{eq:gw_spectrum}. From \cite{Arzoumanian:2020vkk}, the amplitude to explain the data within $2\,\sigma$ is 
\begin{align}
1.8 \times 10^{-15} \lesssim A \lesssim 3.7 \times 10^{-15}\ .
\end{align}
As stated in \cite{Arzoumanian:2020vkk}, the five lowest frequency bins constitute $99.98 \, \%$ of the signal-to-noise contribution, among which the first bin marks the major contribution. The error bar becomes significant already at the third bin. Thus, for the fitting, two bins around
\begin{align}
f_1 \approx 2.5\times 10^{-9}\,{\rm Hz}\ , \quad
f_2 \approx 4.9\times 10^{-9}\,{\rm Hz}
\end{align}
are the most relevant, and we concentrate on the frequency range $f \in [f_1,f_2]$ in the following discussion.
Combining \eqref{OmegaGW_transfer} and \eqref{OmegaGW_NANOGrav}, we can estimate the required GW energy density at the production
\begin{align}
&8.3\times 10^{-5}\left(\frac{g_{s,{\rm gen}}^{4/3}}{g_{*,{\rm gen}}}\right)\left(\frac{f}{f_{\rm yr}}\right)\lesssim \Omega_{\rm GW, gen}
\nonumber \\
& \qquad\qquad\qquad\quad
\lesssim 3.5\times 10^{-4}\left(\frac{g_{s,{\rm gen}}^{4/3}}{g_{*,{\rm gen}}}\right)\left(\frac{f}{f_{\rm yr}}\right)\ .
\label{OmegaGW_needed}
\end{align}
In order for our model to account for signal amplitudes of the NANOGrav observation, we require $\Omega_{\rm GW, \gamma'}$ to be within the range given in Eq.\,\eqref{OmegaGW_needed}, at least at the higher frequency we are interested in, i.e.~$f=f_2$, giving
\begin{align}
\label{OmegaGW_needed_2}
1.3\times 10^{-5} \, g_{*,{\rm gen}}^{1/3}
& \lesssim\Omega_{\rm GW, \gamma'}(c\,p_s)
\lesssim5.4\times 10^{-5} \, g_{*,{\rm gen}}^{1/3} \ .
\end{align}
where $p_s \equiv k_s / a_{\rm gen}$, and we have taken $g_{s,{\rm gen}} = g_{*, \rm gen}$ under the assumption that all the relativistic components are in thermal equilibrium at the time of production.
Here, we have introduced a parameter $c\lesssim 1$ to parameterize the extent by which $f_2$ is lower than the frequency of the GW peak produced by the dark photon. Focusing on the parameter space with $a_{\rm br}\approx a_{\rm tac}$ and taking $a_{\rm gen}=a_{\rm tac}$, the GW spectrum in Eq.~\eqref{eq:gw_spectrum} is reduced to
\begin{align}
&\Omega_{\rm GW, \gamma'}(c\,p_s)\approx 
3\times 10^{-2} \, c\left(\frac{\phi_{\rm osc}}{\Mpl}\right)^4\left(\frac{m_\phi}{H_{\rm osc}}\right)^{5/3}\ .
\end{align}
Using this formula, the condition in Eq.\,\eqref{OmegaGW_needed_2} is reduced to
\begin{align}
\label{eq:condition_2}
&1\lesssim c\,\left(\frac{\phi_{\rm osc}/\Mpl}{0.11}\right)^4\left(\frac{m_\phi/H_{\rm osc}}{3}\right)^{5/3}\lesssim 4 \ ,
\end{align}
where we take $g_{*,{\rm gen}}=10.75$.
We thus gather that, to explain the NANOGrav signal, the axion oscillation amplitude must be close to the Planck scale.

Besides the spectrum amplitude, the spectral behavior needs to be consistent with the NANOGrav observation. As seen in \eqref{eq:gw_spectrum}, the spectral index of our GW is $1$, corresponding to $\gamma = 4$ in \eqref{OmegaGW_NANOGrav}.
The present value of the physical wave number $p_0 $ can be related to the value at the time of production, $p_{\rm gen} $, by $a_{\rm gen} p_{\rm gen} = a_0 p_0 $.
Estimating ratios of the scale factor at different times by those of energy densities, and assuming that the production occurs during the radiation-dominated era, we can relate the value of $p_{\rm gen} $ to the temperature at the production, $T_{\rm gen}$, by
\begin{equation}
    p_{\rm gen} 
    \approx 3.5 \times 10^{-19} \, g_{s, {\rm gen}}^{1/3} \left( \frac{f}{1 \, {\rm yr}^{-1}} \right) T_{\rm gen}
    \; ,
    \label{p_needed}
\end{equation}
To explain the signal frequency, we require the peak frequency is higher than the observed second lowest frequency $f_2=4.9\times 10^{-9}$. This condition is given by $k_s/a_{\rm gen} \gtrsim p_{\rm gen}$, with $k_s / a_{\rm gen}$ found in \eqref{eq:dominant_momentum}, and reduces to
\begin{align}
& \left(\frac{m_\phi}{2.5 \times 10^{-13}\,{\rm eV}}\right)^{1/2}\left(\frac{m_\phi/H_{\rm osc}}{2}\right)^{7/6}\left(\frac{\alpha \phi_{\rm osc}/f_\phi}{30}\right)^{2/3}\gtrsim 1\ .
 \label{eq:condition_1}
\end{align}
This implies that our axion has a small mass around $m_\phi \sim 10^{-13} \, {\rm eV}$ and that the axion starts to oscillate at $T_{\rm osc}\lesssim100\,{\rm MeV}$.

If the axion continues to oscillate coherently, it behaves as matter and dominates the universe soon after the end of the dark photon production, due to its large amplitude. To solve this problem, one possibility is that the axion decays into radiation before it dominates the Universe. However, the quick decay of the axion is difficult due to the shift symmetry of the axion.%
\footnote{An efficiet conversion from the axion to another axion may be achieved through their mass mixing, {\it \`{a} la} Mikheyev-Smirnov-Wolfenstein effect in neutrino oscillations \cite{Kitajima:2014xla,Ho:2018qur}.}
Another possibility is the axion becomes massless before it dominates the universe.
Although this is in a way opposite to a common scenario of symmetry breaking, since the axion's shift symmetry is restored at a later time, this kind of possibility is discussed in~\cite{Barr:2014vva} in the context of the QCD axion. The basic idea is as follows: recall the case of the QCD axion, for which, if there is a massless quark, the $\theta$-parameter becomes unphysical and thus the axion remains massless even after the QCD confinement. We can apply this to e.g.~a hidden QCD sector. 
Let us introduce a vector-like hidden quarks $Q,~\bar Q$ which become massive after a complex scalar field $X$ obtains a non-zero vacuum expectation value (VEV). Then, we consider the axion obtains the mass below the dark QCD confinement temperature.The axion becomes, however, massless again if the VEV of $X$ is changed by $\langle X\rangle\neq 0\to \langle X\rangle=0$ (this inverse-phase transition is already considered in~\cite{Weinberg:1974hy,Langacker:1980kd}.). 
We note that the axion does not disappear even after $\langle X\rangle=0$ if the axion is provided by the other hidden quarks and scalar fields.
Therefore, in this paper, we assume that the axion behaves as radiation soon after the photon production stops. We also discuss the axion abundance searched by the lattice simulations in some previous work in Sec.~\ref{sec:conclusion}.

The abundance of the axion is constrained from the observation of the extra effective neutrino number ${\mit \Delta}N_{\rm eff}$ because the axion behaves as dark radiation after the dark photon production, as discussed in the last paragraph. Assuming the dark sector energy density is dominated by the axion,%
\footnote{If the dark sector temperature is much less than the SM sector, the dark sector thermal bath energy density is negligible.}
the ratio of the dark sector energy density $\rho_{\rm DR, \phi}$ to the total energy density $\rho_{\rm tot}$ at the end of the tachyonic regime ($=$ end of production) is given by
\begin{align}
\label{eq:ds_density}
&\frac{\rho_{\rm DR, \phi}}{\rho_{\rm tot}}\approx\left.\frac{\frac{1}{2}m^2_{\phi}\phi_{\rm osc}^2}{\rho_{\rm tot}}\right|_{a=a_{\rm osc}}\left(\frac{a_{\rm tac}}{a_{\rm osc}}\right)
\approx \frac{2}{3}\left(\frac{\phi_{\rm osc}}{\Mpl}\right)^2\left(\frac{a_{\rm tac}}{a_{\rm osc}}\right)\ ,
\end{align}
where we have identified the starting time of oscillation by $H_{\rm osc} = m_\phi /2$.
Here we have assumed that the axion behaves as radiation right after $a_{\rm tac}$. On the other hand, the dark sector energy density $\rho_{\rm DR}$ at $a=a_{\rm gen}$ is in general written in terms of ${\mit \Delta}N_{\rm eff}$ as~\cite{Nakai:2020oit}
\begin{align}
\label{eq:Neff}
\frac{\rho_{\rm DR}}{\rho_{\rm tot}}=0.07\left(\frac{{\mit \Delta}N_{\rm eff}}{0.5}\right)\left(\frac{g_{s,{\rm gen}}}{g_{s,0}}\right)^{4/3}\left(\frac{g_{*,0}}{g_{*,{\rm gen}}}\right)\ .
\end{align}
The effective number ${\mit \Delta} N_{\rm eff}$ is defined as 
\begin{align}
\rho_{\rm DR}\equiv\frac{7}{8}{\mit \Delta}N_{\rm eff}\left(\frac{4}{11}\right)^{4/3}\frac{2\pi^2}{30}T^4 \; ,
\end{align}
at the recombination time, and thus $T$ is traced back to the value at the time $a_{\rm gen}$ to obtain \eqref{eq:Neff}. 
Using Eq.\,\eqref{eq:ds_density} and Eq.\,\eqref{eq:Neff}, we obtain the relation,
\begin{align}
\left(\frac{\phi_{\rm osc}/M_{\rm PL}}{0.11}\right)^2\left(\frac{a_{\rm tac}/a_{\rm osc}}{10}\right) \approx 
\left(\frac{{\mit \Delta}N_{\rm eff}}{0.5}\right)\left(\frac{g_{*,{\rm gen}}}{10.75}\right)^{1/3}\ .
\label{eq:condition_3}
\end{align} 
The observational requirement is ${\mit \Delta}N_{\rm eff}\lesssim 0.7$ from $N_{\rm eff}=3.27\pm0.15~~~(68\%\,{\rm C.L.})$~\cite{Aghanim:2018eyx,Riess:2018uxu}.
The Hubble tension is reconciled by ${\mit \Delta}N_{\rm eff}\sim 0.5$~\cite{Bernal:2016gxb,Aghanim:2018eyx,Blinov:2020hmc}, and thus the parameter values that account for the NANOGrav observation in our model may simultaneously serve a mechanism to alleviate the tension.
We note, however, that, as mentioned in the paragraph below \eqref{eq:atac}, the true value of $a_{\rm tac} / a_{\rm osc}$ might be larger than the analytically obtained one $\approx 10$. In such cases, $g_{*,{\rm gen}}$ would necessarily take a larger value to satisfy the bound on $\Delta N_{\rm eff}$, or more preferably to account for the Hubble tension. An accurate evaluation of $a_{\rm tac}$ requires to take into account the effects of back reaction, which is beyond the validity range of our analytical calculation, and we would like to leave this consideration to future studies.

In summary, we obtain three conditions to explain the NANOGrav signal, and possibly the tension in the determinations of the Hubble constant. From Eq.\,\eqref{eq:condition_2}, Eq.\,\eqref{eq:condition_1}, and Eq.\,\eqref{eq:condition_3}. the typical parameter values are 
\begin{align}
m_\phi\sim 10^{-13}\,{\rm eV}\ ,~\phi_{\rm osc} \sim0.1 \Mpl \ ,~ \frac{\alpha\,\phi_{\rm osc}}{f_\phi} \sim30\ .
\end{align}
Note that we focus on the parameter values with which the resonance stops at $a_{\rm tac}\approx a_{\rm br} $, where GW are maximally produced. We also note that the constraint from the superradiance~\cite{Arvanitaki:2009fg,Arvanitaki:2010sy,Arvanitaki:2014wva} is avoided, since we assume the axion has been massless since the end of the production until present.

%%%%
\begin{figure}
\centering
  \includegraphics[width=1\linewidth]{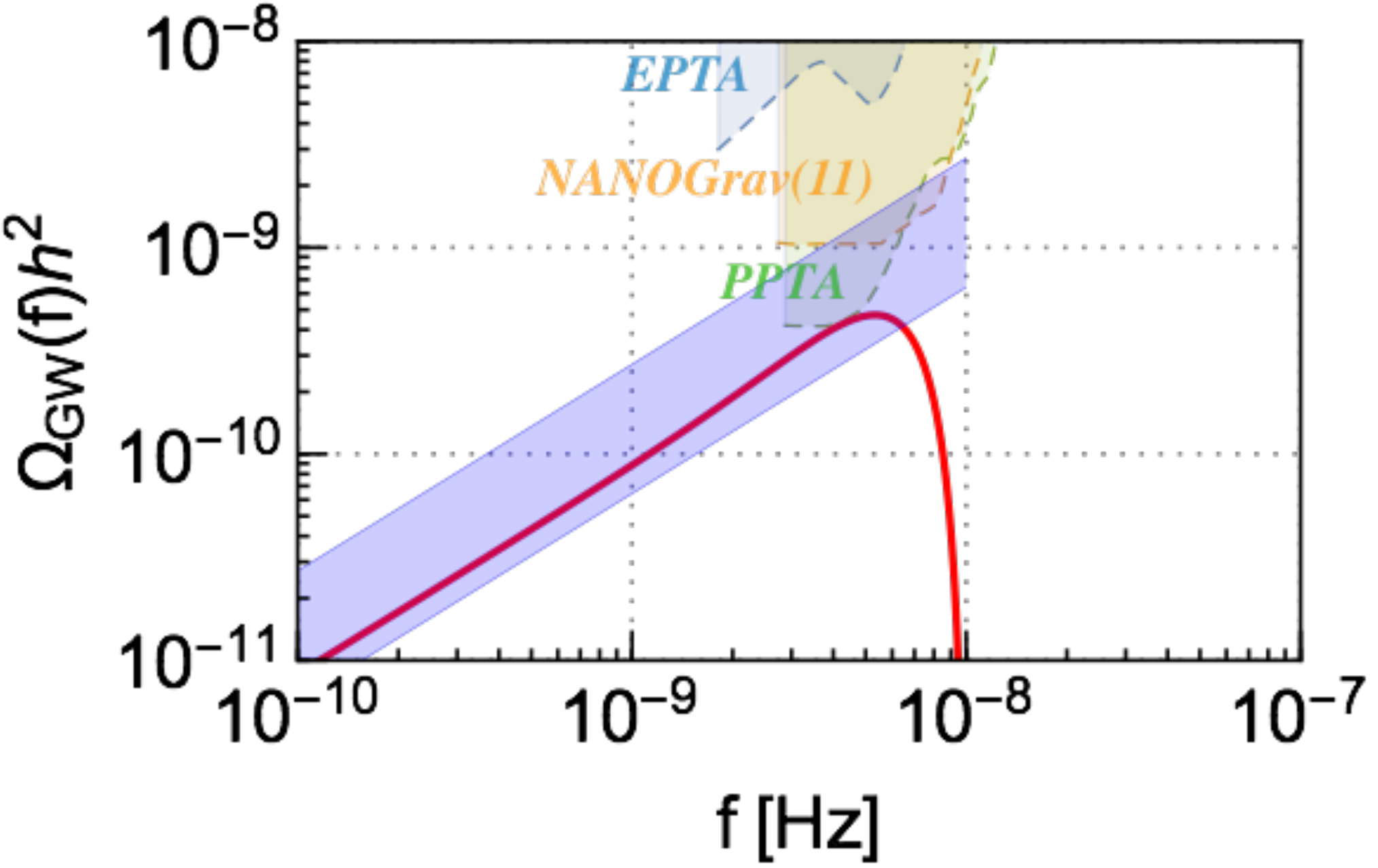}
   \caption{Comparison of the GW spectrum originated from the produced photon and the NANOGrav power-low model.
The red line denotes the GW spectrum of the photon for $\phi_{\rm osc}=0.12\,M_{\rm PL}$ and $m_\phi=10^{-12.5}\,{\rm eV}$. The blue shaded region corresponds to the observed NANOGrav GW amplitude modeled by power-low model with $\gamma = 4$ within $2\sigma$. A cutoff is placed around $10^{-8} \, {\rm Hz}$, reflecting large error bars in the NANOGrav result above this frequency range.}
\label{fig:spectrum}
\end{figure}
%%%%  

In Fig.~\ref{fig:spectrum}, we show an example spectrum where the above parameter conditions are satisfied. The red line corresponds to the GW spectrum produced by axion-photon resonance at the parameter of $\phi_{\rm osc}=0.12\,M_{\rm PL}$ and $m_\phi=10^{-12.5}\,{\rm eV}$. The blue shaded region is favored by a power-law model with $\gamma=4$ within $2\sigma$.
We place a cutoff for the blue region around $10^{-8} \, {\rm Hz}$, reflecting large error bars in the NANOGrav data above this frequency range.
We find a good agreement with the power-law model and the GW produced by the axion-photon resonance.

\section{Discussion and conclusion}
\label{sec:conclusion}
The dynamics of axion-like fields and gauge fields in the presence of their interaction has been an active area of research. Violent production of the gauge quanta due to the resonance and tachyonic growth induced by the coherent oscillation of the axion entails rich phenomenological signatures. Such produced quanta develop large quadrupole moments and act as an efficient source of gravitational waves.
In this paper, we have employed this production mechanism of a $U(1)$ gauge field present beyond the Standard Model, which we call dark photon, and computed the resulting spectrum of stochastic GW signals, with the recent pulsar timing observation NANOGrav as the main observational target.

The production is particularly efficient for a large coupling, the case we focus on in this paper. In the course of a single oscillation of the axion, each mode of the dark photon goes through $4$ stages: damped oscillation by positive $\omega^2$, momentary violation of adiabaticity condition, tachyonic behavior due to negative $\omega^2 $, and another short period of adiabaticity violation. Solving each stage separately, and connecting the solutions at the overlapping regions, we obtain the analytical formula that well approximates the dark photon behavior at all time during production.
Using it, we then adopt the Green function method to compute the contribution to the GW spectrum.
In order for this GW to account for the reported NANOGrav result \cite{Arzoumanian:2020vkk}, especially its fist few frequency bins that dominate the overall signal-to-noise ratio, we find the required parameter values should be $m_\phi \sim 10^{-13} \, {\rm eV} $, $\phi_{\rm osc} \sim 0.1 \, \Mpl $ and $f_\phi / \alpha \sim 10^{16} \, {\rm GeV}$, yielding our main result in this work.

The production in our scenario necessarily occurs during the radiation-dominate universe. If the axion continued to oscillate after the dark photon production ends, its density would increase relative to the total background density and would soon over-dominate the universe, for the parameter values mentioned above. In Sec.~\ref{sec:GW}, to avoid this problem occurring, we have discussed an inverse-type phase transition that recovers the axion massless after the temperature drops below some critical value. We here admit a tuning so that such a transition in the dark sector, which contains the axion of our interest, takes place soon after the production ceases.

There is, however, an alternative scenario that may suppress the axion abundance without an additional ingredient, though more computationally involved.
In this paper, we have focused on the case in which the back reaction effect is under control. Once it becomes important, on the other hand, a significant fraction of the axion energy could be transferred to the dark photon.
Ref.~\cite{Agrawal:2017eqm} numerically solves the axion-dark photon system with the initial condition of $\phi_{\rm osc} = f_\phi $ for $f_\phi = 10^{16 - 17} \, {\rm GeV}$ and $\alpha = 20 - 60 $. Their calculations exhibit an exponential suppression of the axion energy density even after the energy density of the dark photon becomes comparable to that of the axion. Eventually the axion energy density settles down at the value that can explain the current dark matter density.
In Ref.~\cite{Kitajima:2017peg}, however, lattice simulation is performed and does not confirm such a significant suppression even for similar axion parameters. The latter simulation even exhibits an enhancement of the axion density for $\alpha \gtrsim 200$ due to a considerable friction by the produced dark photon, as compared to the case of negligible interaction $\alpha = 0$.
While this discrepancy in the dynamics when the energy densities of the two components become comparable is yet to be understood and is beyond the scope of our current study, there appears to exist a parameter space in which the dark photon absorbs a significant fraction of the axion's initial energy.
In such a case, the axion density may sufficiently decrease to the level subdominant to the dark matter density, or possibly just to a level that can fully account for the whole dark-matter abundance. This is certainly an intriguing and attractive possibility, which, however, requires a consistent treatment of the back reaction from the produced dark photon onto the axion dynamics, and thus we leave it to our future investigations.

The current report of stochastic GW background signal by NANOGrav shows null evidence for quadrupolar spatial correlations and may suffer unincluded and/or unknown systematics. Further analyses of the data and observations by other pulsar-timing missions, such as the Parkes Pulsar Timing Array (PPTA) \cite{Manchester:2012za,Kerr:2020qdo} and European Pulsar Timing Array (EPTA) \cite{Kramer:2013kea,Lentati:2015qwp}, are mandatory to confirm or refute the true identity of the signal. Yet, if it were to be confirmed, that would certainly provide important implications about the physics in the early universe. We have demonstrated one stimulating example, connecting the physics with axion-like fields beyond the Standard Model and the ongoing GW searches.
We will extend the study of the ALF-gauge field dynamics for broader applications and show the details of our analytical calculations in our upcoming publication.

\vspace{5mm}
\noindent
{\bf Note added}: At the final stage of preparation of our paper, Ref.~\cite{Ratzinger:2020koh} was posted, which, based on \cite{Machado:2018nqk,Machado:2019xuc}, aims at GW generation of dark photon production by the motion of axion-like fields, similar to our consideration in this paper. The major difference is that, while their study is based on numerical computations, our calculations are analytical with a clear validity range, consistent with the result in \cite{Salehian:2020dsf}. Our result is essentially compatible with \cite{Ratzinger:2020koh} in terms of the resultant parameter window for the considered model, albeit the different approaches. As we discussed in Sec.~\ref{sec:GW}, however, the axion-like field in this model would easily over-dominate the universe, unless rendered harmless. In this paper, we have explicitly discussed a possible way out to avoid such a pathological scenario.

\vspace{5mm}
\noindent
{\bf Acknowledgment}:
M.S. thanks F.~Takahashi for several useful comments and pointing out \cite{Kitajima:2014xla,Ho:2018qur} in a private communication. M.S. also thanks M.~Yamada, Y.~Nakai for discussions.
R.N.~is grateful to B.~Cyr for useful discussions on the gauge field production.

%%%%%%%%%%%%%%%%%%%%%%%%%%%%%%%%%%%%%%%%%%%%%%%%%%%
%%%%%%%%%%%%%%%%%%%%%%%%%%%%%%%%%%%%%%%%%%%%%%%%%%%
\bibliographystyle{apsrev4-1}
\bibliography{axiongauge}

\end{document}